# First estimations of Gravity Wave Potential Energy in the Martian thermosphere: An analysis using MAVEN NGIMS data

G. Manju [1] and Mridula N. [1]

E-mail: manju_spl@vssc.gov.in

List of institutions

1 Space Physics Laboratory, Vikram Sarabhai Space Centre,

ISRO P.O, Thiruvananthapuram, India. pin 695022. Ph: +91-471-2562597 Fax: +91-471-2706535

Abstract

First estimations of Gravity Wave Potential Energy (GWPE) for the Martian thermosphere are reported herein using the height profile of $CO_2$ density derived temperature fluctuations for different Martian seasons during the 33rd Martian year. Explicit diurnal evolution of GWPE (for 52º to 73º latitude bin) with a post sunset maximum is delineated for summer. The higher values of GWPE are observed during morning, compared to post mid-night (35º to 55º latitude bin) for summer. As the latitude increases from 16º to 45º, GWPE (1-4 LT bin) is found to be nearly doubled for summer. Further, GWPE estimates in autumn are 6 times higher during night compared to day (-45º to -72º latitude bin) and daytime (-53º to -72º latitude bin) GWPE is much lower in autumn compared to spring for all longitudes. Overall, from the available data autumn (with respect to northern hemisphere) daytime periods appear better suited for aero-braking operations of Martian landing missions.

## 1. Introduction

Gravity waves are perturbations propagating in stratified fluids, with buoyancy (gravity) as the restoring force. As a result, periodic oscillations are set up in the constituent fluids in the atmosphere (both in space and time) resulting in variability of temperature, pressure, density etc. [Gossard and Hooke (1975)]. The impact of gravity waves in the terrestrial atmosphere/thermosphere system was brought out by several studies in the past [Vadas and Fritts (2001); Vadas and Fritts (2006); Sreeja et al., (2009); Manju and Aswathy (2017); Aswathy and Manju (2018); Mridula and Pant (2017)].

Important observations pertaining to the amplitude as well as vertical and horizontal wavelengths of gravity waves have been brought out in the past. Fritts, Wong and Tolson (2006) showed that gravity waves in the Martian thermosphere have large amplitudes of the order of 5 percent to 50 percent. Ando et al., (2012) using MGS data showed the vertical wave number spectra of gravity waves in the Martian atmosphere. Yigit et al., (2015a) showed that the relative density perturbations between 180 km and 220 km were around 20 percent to 40 percent and showed significant local time, latitude and altitude variations. Their modeling study also showed that gravity waves could propagate from lower atmosphere to thermosphere. Terada et al., (2017) observed that the average amplitude of the perturbations is 10 percent on the dayside and 20 percent on the night side. Zurek et al., (2017) demonstrated that neutral density pattern exhibits seasonal and solar zenith angle variations. Following this, Siddle et al., (2019) observed that with increasing solar zenith angles, the gravity wave amplitudes also showed enhancement. Williamson et al., (2019) showed that gravity wave dissipation affected the composition of the exosphere and led to an enhancement in $O/CO_2$ ratio. Creasy, Forbes and Keating (2006b) reported the global and seasonal distribution of gravity wave activity in Martian atmosphere using MGS radio occultation data.

As discussed above, in addition to determining the spatial and temporal variability of atmospheric state parameters like density, temperature etc., the vertically propagating gravity waves originating in the lower atmospheric regions are capable of modulating the neutral motion as such. Gravity waves have the potential to slow down or reverse the mean flow in the upper thermosphere as they impart their momentum to the mean flow. Based on the local conditions, the atmosphere undergoes either warming or cooling [Medvedev et al., (2015)] thereby altering the Martian energy budget [Kuroda et al., (2015)] resulting in altered circulation and transport processes. Gravity wave associated cooling leads to cloud formation [Medvedev et al., (2015)] which is an important concern for the aerobraking phases of mars missions.

In addition, structures which appear in the neutral atmosphere due to gravity waves and other dynamical processes can produce similar deviations in plasma density distributions also [Mayyasi et al., (2019)] thereby acting as primary driving force for producing ionospheric variability like plasma turbulence. Further the heating of the upper atmosphere by gravity waves is shown to contribute to atmospheric escape via the jeans escape by Walterscheid, Hickey and Schubert

(2013) in modeling study. Hence in order to have a handle on the loss processes, an understanding of the basic evolution of the gravity wave dynamics is very essential.

From the above discussion it is clear that the distribution of gravity wave potential energies in the Martian atmosphere/thermosphere has not been done so far. The energy per unit mass, E, is the most important parameter which gives a measure of the gravity wave activity at any height in the atmosphere (Wilson Chanin and Hauchecorne (1991a), Wilson Chanin and Hauchecorne (1991b)). Hence the present analysis attempts to bring out a clear picture of the spatio temporal evolution of the longitudinal distribution in gravity wave potential energy for all the available seasons of the 33rd Martian year.

2 DATA, METHODOLOGY AND OBSERVATIONS

In the present study, the data obtained from the NGIMS instrument aboard MAVEN, a quadrupole mass spectrometer capable of measuring atmospheric densities at altitudes below 500 km above the areoid is used [Benna et al., (2015); Benna and Lyness (2014)]. The altitude range of 160 km to 220 km is considered for the present study, as at higher altitudes the densities decrease and the perturbation signatures diminish [ England et al., (2017)]. The spacecraft is in an eccentric 4.5 hour orbit and has an apoapsis around 6220 km and periapsis around 150 km [Jakosky (2015)]. The NGIMS neutral density data for the years 2015 to 2017 (Level 2, version 07, revision 01) is obtained from the link given below. https://atmos.nmsu.edu/PDS/data/PDS4/MAVEN/ngims_bundle/l2/

The data used spans over a longitude region of -180o to 180o, latitude region of +/-90o and local time range of 0-24 hr. All seasons are referred with respect to northern hemisphere. An analysis is carried out to ensure that the vertical variations are dominantly modulating the observations. Based on these calculations, the vertical variation in density is of the order of 95 percent considering the 160-220 km altitude region, while the horizontal variability within the pass is found to be < 20 percent (considering the data used for the study). Hence, we are essentially sampling the vertical propagation characteristics using the data.

An analysis is carried out to ensure that the vertical variations are dominantly modulating the observations. Based on these calculations, the vertical variation in density is of the order of 95 percent considering the 160-220 km altitude region, while the horizontal variability within the pass is found to be < 20 percent (considering the data used for the study). Hence we are essentially sampling the vertical propagation characteristics using the data.

**2.1 Estimation of GWPE**

A brief description of the technique used to derive atmospheric pressure and temperature is given in the following paragraphs [Snowden et al., (2013); Mueller Wodarg et al., (2006); Gubenko, Kirillovich and Pavelyev (2015)]. Accordingly, the altitudinal density profiles of $CO_2$ are obtained from NGIMS. Thereafter, the pressure is calculated from the $CO_2$ density by integrating the hydrostatic equilibrium equation for $CO_2$ downward from upper boundary to the given altitude to obtain the local partial pressure of $CO_2$.

$$P = P_0 + G.M.m \int_r^{r_0} N(r)\frac{dr}{r^2} \qquad (1)$$

Where, G is the universal gravitational constant, M is the mass of Mars, m is the molecular weight of the $CO_2$, N(r) is the density and P(r) is the pressure at distance r from martian centre and $r_0$ is the upper altitudinal boundary of the density profile. The quantity $P_0$ is an integration constant that represents the value of pressure at $r_0$, the upper boundary. $P_0$ is determined by fitting the top three points in the density range to a model for a hydrostatic distribution with a temperature gradient equal to constant times the adiabatic gradient. Taking the derivative of the ideal gas law gives

$$\frac{1}{P}.\frac{dP}{dr} = \frac{1}{T}.\frac{dT}{dr} + \frac{1}{N}.\frac{dN}{dr} \qquad (2)$$

By using hydrostatic equilibrium to replace dP/dr and setting the temperature gradient to constant times the adiabat α we get,

$$\frac{dT}{dr} = -\alpha.\frac{g}{C_p} = -\alpha.\frac{GM}{r^2.C_p} \qquad (3)$$

g is equal to 3.711m/s2 and $C_p$ for $CO_2$ is 0.844 KJ/kgK

$$\frac{dlogN}{dr} = \frac{GM}{T_0.r^2}\left(\frac{\alpha}{C_p} - \frac{1}{R}\right) \qquad (4)$$

where R is the specific gas constant and α =0 [Snowden et al., (2013); Mueller Wodarg et al., (2006)]. Thus, the value of $T_0$ is determined from equation 4. $P_0$ is calculated from the value of $T_0$ and the measured density using the ideal gas law given by equation 5.

P=N.K.T  (5)

Combining the calculated pressure $P_0$ using equation 1, pressure P for the entire altitude range is estimated. From this, with the measured density and the ideal gas law, the temperature for each altitude is estimated as given by equation 6.

T=P/NK  (6)

where K is Boltzmann's constant. From the temperature profiles obtained, the fluctuations are estimated and the profiles of Brunt-Vaisala frequency and Gravity wave potential Energy (GWPE) are estimated by using the equations 7 and 8 respectively.

$$N_b^2(z) = \frac{g(z)}{T_0(z)} \left[ \frac{\partial T_0(z)}{\partial z} + \frac{g(z)}{C_p} \right] \quad (7)$$

$$E_p(z) = \frac{1}{2} \cdot \left( \frac{g(z)}{N_b(z)} \right)^2 \cdot \left( \frac{T(z)'}{T_0(z)} \right)^2 \quad (8)$$

Where, $g(z)$ is the acceleration due to gravity, $T_0$ is the mean temperature at an altitude z, $T(z)'$ is the temperature fluctuation of the instantaneous temperature profile about the mean temperature $T_0$ and $N_b(z)$ is the Brunt-Vaisala frequency.

The various parameters estimated from the NGIMS density data on 20 September 2015, are represented in figure 1 panels a to f.

The density profile represented in panel a of figure 1 is obtained from NGIMS data directly. From this profile, by applying the above methodology, pressure is determined and the same is depicted in panel b of figure 1. From the altitudinal profile of pressure, temperature between 160 km and 220 km is calculated (panel c).

The background temperature is estimated by using a thirty-point adjacent averaging to include all the major vertical wavelengths of perturbations present [Gardner et al., (2011)] and is represented by the brown line in panel c. The implementation of the 30-point adjacent averaging, ensures that the contamination from thermal tides is excluded. Other workers have also reported thermal tidal wavelengths of this order [ Hinson and Wilson (2004), Ando et al., (2012)).

The temperature fluctuations are estimated by subtracting the actual profile from the smoothed background temperature profile. Fluctuations are plotted as a function of altitude in panel d. The profiles of Brunt vaisala frequency and energy are estimated by using the equations 7 and 8 and are plotted in panels e and f respectively. For a given orbit, the mean value of GWPE is estimated. The mean GWPEs for all the available orbits for each martian season are extracted to obtain the longitudinal distribution of the same. The seasons are classified based on inputs obtained from www.planetary.org/explorespace-topics/mars/calender.html. The seasons are Spring (July 2015 to December 2015), Summer (January 2016 to June 2016), Autumn (July 2016 to November 2016) and Winter (December 2016 to May2017).The longitude bands 180 to -120, -120 to -60, 0 to 60, 60 to 120 and 120 to 180 are classified as bins 1 to 6 respectively. The seasonal patterns for the different time sectors have been derived as per the data availability in the respective hemispheres for those time sectors.

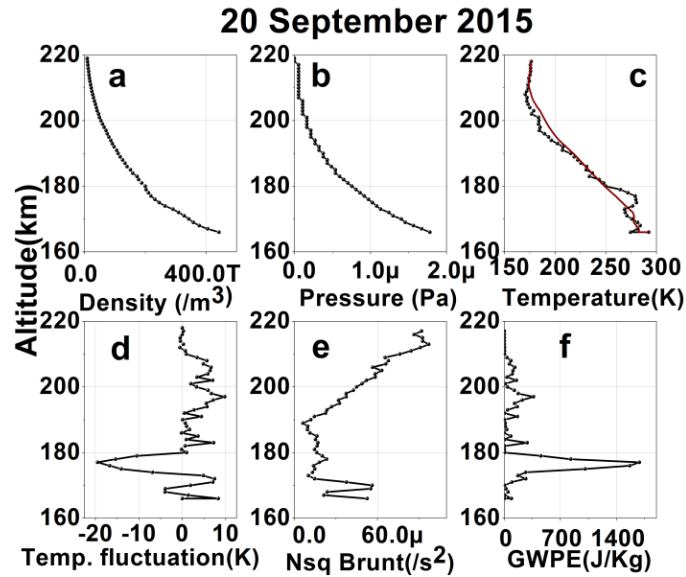

**Figure 1.** The altitude profiles of density, pressure, temperature, temperature fluctuation, $N^2$ and GWPE for 20 September 2015, are represented in figure Figure 1 a, panels a to f.

**2.2 Wavelet analysis of Temperature fluctuations on 20 September 2015**

Figure 2 depicts the wavelet spectrum for the temperature fluctuations obtained on 20 September 2015. It is seen that the scale sizes in the band of 10 km to 20 km dominate in the entire altitude region, with other shorter scale sizes around 5 km and below also manifesting. In view of the dominance of these particular scale sizes in the wavelet spectrum, the scale sizes within 30 km have been extracted by applying 30-point running mean to the raw temperature profiles and then removing the same from the raw profiles.

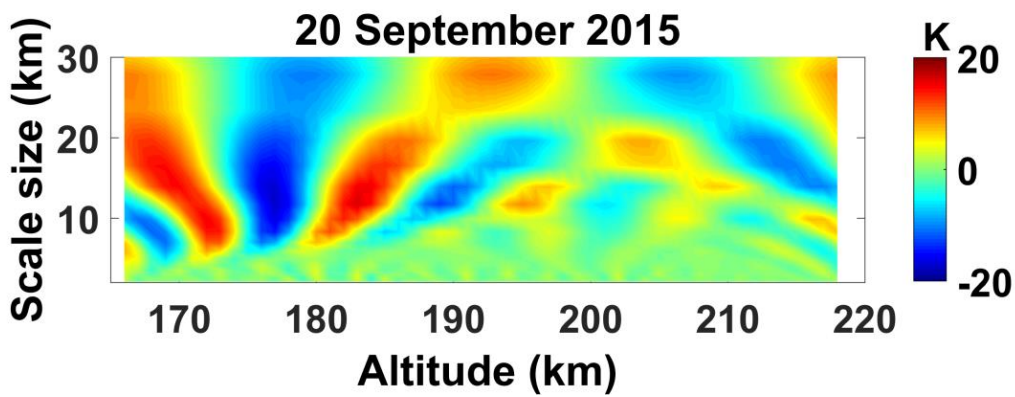

**Figure 2.** The wavelet spectrum for the temperature fluctuations obtained on 20 September 2015.

Since the dominant vertical wavelength is much below the thermal tidal vertical wavelength domain of above 30 km it is reasonable to attribute these fluctuations to gravity waves ([ Hinson and Wilson (2004)). Creasy et al., 2006a, reported gravity waves with vertical wavelengths in the range of 8 to 15 km for the martian altitude regions below 30 km. Siddle et al., (2019) observed gravity waves with vertical wavelengths of 10 to 30 km for the martian thermospheric altitudes of 150 km to 180 km. Therefore, the present study is in broad agreement with the above mentioned results.

## 2.3 Estimation of gravity wave parameters of the sample day of 20 September 2015.

The gravity wave parameters have been estimated for the sample day of 20 September 2015 based on the expressions given by Dornbrack, Gisinger and Kaifler (2006).The horizontal wave number, k = $2\pi/\lambda_H$ and vertical wavenumber, m =$2\pi/\lambda_V$ where $\lambda_H$ and $\lambda_V$ are the horizontal and vertical wavelengths respectively. The vertical wavelength is derived from the wavelet analysis given in section 3.1. Observed vertical wavelength in the present study is 15 km. Creasy,Forbes and Hinson (2006a) revealed gravity wave horizontal wavelengths of 100 km to 300 km in the martian thermospheric altitude region of 100 km to 170 km. Assuming the horizontal wavelengths of 100 km and 300 km, the estimated periodicities fall in the range of 1.86 to 5.52 hours. The corresponding phase velocities are 15.1 m/s and 14.9 m/s respectively. Evidently these periodicities correspond to that of gravity waves and are lower than the tidal periodicites of >6 hours.

## 2.4 Estimation of Gravity Wave Potential Energy (GWPE) for Martian year 33

Gravity Wave Potential Energy (GWPE) for four local times of the 6 longitude bins for the Northern latitudes of around 52 to 73 $^0$ is depicted in figure 3 as a function of altitude (panel a to d). Assuming that the variability within 200 is not so significant, we can obtain an equivalent LT variation of summer. 15 LT shows minimum GWPE across all LT sectors and altitudes. Maximum GWPE (around 4000J/kg) is observed during 19.5 LT, with the longitude bin of 4 to 6 showing large GWPE in almost the entire altitude range. Bins 1 to 3 show enhanced GWPE in a narrow range of altitudes between 180 km and 190 km. During 23.2 LT, the values show a secondary maximum around 180 km (around 3500 J/kg). The spatial distribution of GWPE values are much more limited for all other time sectors compared to 19.5 LT. Overall, it is evident from the figure that there is almost 100 percent change in GWPE values between the maximum at 19.5 LT and the minimum at 15 LT.

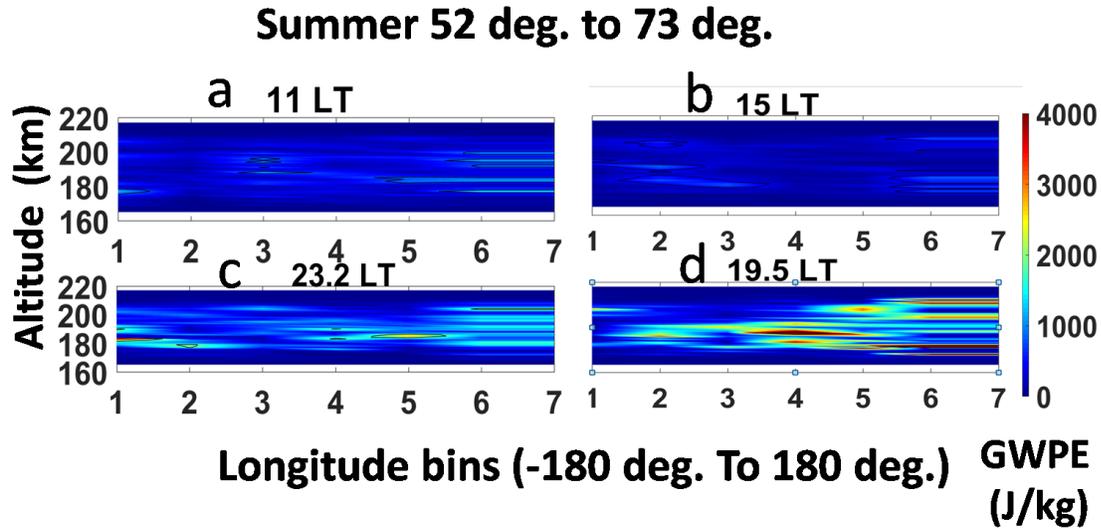

**Figure 3**. Gravity Wave Potential Energy (GWPE) for four local times of the 6 longitude bins for the Northern latitudes bin of $52^0$ to $73^0$, as a function of altitude (panel a to d).

The altitudinal mean of GWPE for each longitude bin is extracted from each of the above 4 LT bins to obtain the temporal evolution of GWPE for different longitude bins as depicted in Figure 4. This figure clearly shows the diurnal evolution of GWPE for $52^0$ to $73^0$ North latitude bin for summer. The important features to be noted are the post sunset maximum in GWPE at 19 LT followed by those at 11 LT, ~23.2 LT and ~15 LT in that order. For a given LT bin the GWPE maximizes at the 120-$180^0$ longitude bin (bin 6).

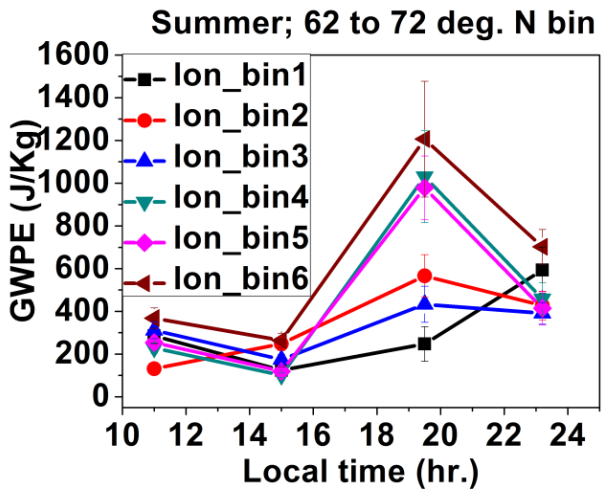

Figure 4. The temporal evolution of GWPE for different longitude bins

Figure 5 panel a depicts the longitudinal pattern of GWPE in the 35-$55^0$ latitude bin for 2 local time bins of summer. The GWPE values are in general higher for 9 LT bin compared to 2.5 LT bin. Figure 5 panel b depicts the longitudinal pattern of GWPE in the -45 to -72$^0$ latitude bin for 14.9 local time bin of autumn.

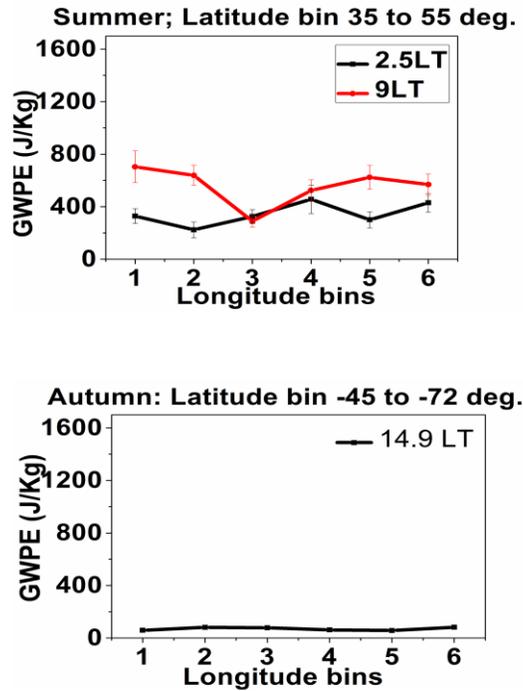

**Figure 5**. a) The longitudinal pattern of GWPE in the 35-55 $^0$ latitude bin for 2 local time bins of summer. b) The longitudinal pattern of GWPE in the -45 to -72 $^0$ latitude bin for 14.9 local time bin of Autumn.

Figure 6 shows the longitudinal variation of GWPE for the 1-4 LT bin for the 450 and 160 north latitude regions of summer. These observations for the $45^0$ and $16^0$ latitude regions show the enhancement (by ~2 times) of GWPE as the latitude increased from $16^0$ to $45^0$

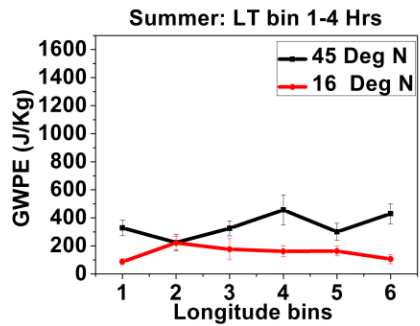

**Figure 6.** The longitudinal variation of GWPE for the 1-4 LT bin for the $45^0$ and $16^0$ latitude regions of summer

Figure 7 panel a illustrates the longitudinal variation of GWPE for the 15-19 LT bin of the -53 to -72$^0$ latitude bin; while Figure 7 panel b depicts the same for the 19-20.9 LT bin of the -40 to -71$^0$ latitude bin; both corresponding to spring and autumn seasons. For the day LT bins there is clear decrease in the GWPE in autumn compared to spring for all longitudinal regions. In the case of the spring night 19-20.9 LT, enhancement in GWPE is observed for the longitude region -120$^0$ to 0$^0$ while similar enhancement is not observed for the 0-120$^0$ region in comparison with the day time behavior. It is observed that the autumn night time GWPE shows the expected night time increase compared to the corresponding day time pattern for all longitude bins. The other notable point is that there is large zonal asymmetry in GWPE during day time in spring while such an asymmetry is not manifesting during night time.

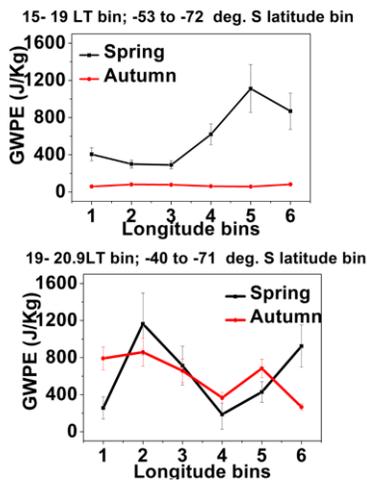

**Figure 7**. a) The longitudinal variation of GWPE for the 15-19 LT bin of the -53 to -72 $^0$ latitude bin for spring and autumn; b) The same for the 19-20.9 LT bin of the -40 to -71 $^0$ latitude bin for spring and autumn

3. Discussions

The distribution of gravity wave potential energies in the Martian atmosphere/thermosphere has not been published so far although the global picture of the gravity wave amplitude variability in the thermosphere region, for different solar zenith angles as well as during different seasons have been obtained [Siddle et al., (2019)]. This work attempts to examine the spatio-temporal variability of GWPE in the Martian thermosphere. The fluctuations in temperature have been extracted and subjected to wavelet analysis to delineate the vertical wavelength of the same. It is observed that the dominant vertical wavelength of ~15 km is well below the vertical wave length domain of thermal tides. Thereafter, the gravity wave characteristics have also been analyzed. For the horizontal wavelength range of 100 km - 300 km corresponding to the observed vertical wavelength of ~15 km, the estimated periodicities fall in the range of 1.86 to 5.52 hours and the corresponding phase velocities are 15.1 m/s and 14.9 m/s respectively. The analysis reveals that the periods are much below tidal periodicities thus confirming the gravity wave origin of the fluctuations.

The spatio-temporal evolution of GWPE for the available data sets has been obtained. Clear diurnal evolution of GWPE for $52^0$ to $73^0$ North latitude bin of summer is brought out in the present study (Figure 3 and Figure 4). The important features that emerged are the post sunset maximum in GWPE at 19 LT with a minimum manifesting at 15 LT. This behavior is very similar to that observed on Earth wherein the GW amplitude is maximum during post sunset hours and low during day time. The role of the sunset terminator in producing enhanced GW activity has been suggested for the above diurnal pattern on Earth [Manju and Aswathy (2017)]. Liu, Li and Jin., (2019) reported a diurnal cycle in ambient Martian thermospheric neutral densities for all seasons with a maximum during day time (15 LT) and a minimum in the close to mid-night period. The maximum in Martian thermospheric GWPE occurring during post sunset hours also points towards the combined role of sunset terminator and diurnal neutral density pattern in producing the observed behavior. The higher values of GWPE during the post sunset /night time period compared to the low values during noon to post noon hours is a notable observation in view of the potential hazardous effect of GWs on aero braking phases of Mars missions.

The longitudinal pattern of GWPE in the 35-55$^0$ latitude bin for 2 local time bins of summer are examined (Figure 5a). The higher values of GWPE during 9 LT compared to 2.5 LT is possibly due to the effect of the sun rise terminator, which probably acts as a source of enhanced gravity wave activity, akin to the sun set terminator during post sunset hours. The longitudinal pattern of GWPE in the -45 to -72$^0$ latitude bin for 14.9 local time bin of autumn have also been examined(Figure 5b). It is seen that GWPE in autumn (daytime) is substantially lower compared to other available seasons for all available latitude sectors (for example Summer).Liu, Li and Jin.,(2019)have reported that the strongest diurnal neutral density cycle amplitude prevails in autumn with a distinct afternoon peak (LST = 15 h), and a rapid drop during the night. They have attributed these trends to comparatively rapid cooling processes during the night in autumn. Accordingly, for the higher neutral density prevalent at daytime of autumn, the quenching effects on GWs are higher thereby lowering gravity wave activity. Such a scenario explains the lower GWPE reported herein.

The latitudinal variation of the longitudinal distribution of GWPE for the 1-4 LT bin of summer is examined in this study (Figure 6). An enhancement (by as much as a factor of 2) of GWPE is observed as the latitude increases from $16^0$ to $45^0$. Yigit et al., (2015a) reported that the amplitudes of the GW induced density fluctuations are somewhat lower in the $60^0$ to $65^0$ latitude region (20 percent) relative density perturbation) compared with the region above $65^0$N latitude giving indication of a latitudinal increase of GW amplitudes. The present observation seems to show similar behaviour.

The longitudinal variation of GWPE for the 15-19 LT bin of the -53- to -72$^0$ latitude bin; and the same for the 19-20.9 LT bin of the -40 to -71$^0$ latitude bin; are both examined corresponding to spring and autumn seasons (Figure 7). For the day LT bins there is clear decrease in the GWPE in autumn compared to spring for all longitudinal regions. This is attributed to the large neutral densities prevailing in autumn leading to greater quenching of GW activity. The night time increase in GWPE is observed for the longitude region $-120^0$ to $0^0$ while the similar enhancement is not observed for the $0-120^0$ region. In view of the prevailing low neutral densities during night time, the absence of enhancement in GWPE in the $0-120^0$ bin needs to be understood. Further, the spring night time pattern does not reflect the expected distinct modulation by neutral density. Another observation is that there is a large zonal asymmetry in GWPE during daytime of spring while such an asymmetry is not observed during night time.

Precise aerobraking operations are extremely important for the safety of mars lander missions. In this context, considering all the available longitudes and seasons, it is concluded that day time is preferable for aerobraking operations, compared to night time. Further, it is observed that GWPE values are higher in spring compared to autumn, implying that safe aerobraking operations are plausible during the latter. Overall, the present study shows the potential of GWPE studies in better fine tuning the aero braking phases for Mars lander missions. Similar studies with data availability in more spatial and temporal domains need to be planned in future to arrive at fool proof landing scenarios.

## 4. Conclusions

Gravity Wave Potential Energy (GWPE) estimations for the Martian thermosphere are reported herein for the first time. The analysis conducted using the MAVEN data for the $33^{rd}$ Martian year, reveals the following aspects.

i)Unambiguous diurnal evolution of GWPE (for $52^0$ to $73^0$ North latitude bin) with a post sunset maximum is delineated for summer.

ii) The higher values of GWPE, during morning hours, compared to post mid- night period in the $35^0$ to $55^0$ latitude bin for northern hemisphere summer.

iii) GWPE in autumn (day time) is substantially lower compared to other available seasons in different latitude sectors.

iv) The analysis of the latitudinal variation of GWPE for the 1-4 LT bin of northern hemisphere summer reveals near doubling of GWPE as the latitude increases from $16^0$ to $45^0$.

v) For the day LT bins (in the $-53^0$ to $-72^0$ latitude bin) there is a clear decrease in the GWPE in autumn compared to spring for all longitudinal regions.

vi) A comparison of the day and night time observations in spring demonstrates an increase in GWPE for the longitude region $-120^0$ to $0^0$ while the similar enhancement is not observed for the $0^0$ -$120^0$ region.

vii) Autumn daytime periods are found to be more suitable for aerobraking operations.

**Acknowledgements**

The authors acknowledge Indian Space Research Organization for supporting this work. The NGIMS neutral density data for the year 2015 to 2017 (Level 2, version 07, revision 01) is obtained from the link given below.

https://atmos.nmsu.edu/PDS/data/PDS4/MAVEN/ngims_bundle/l2/

The authors thank the MAVEN team for the data. The seasons are classified based on inputs obtained from www.planetary.org/explorespace-topics/mars/calender.html

**Data Availability**

The data underlying this article are available in https://atmos.nmsu.edu/PDS/data/PDS4/MAVEN/ngims\_bundle/l2/. The data sets were derived from sources in public domain https://atmos.nmsu.edu/PDS/data/PDS4/MAVEN/ngims\_bundle/l2/

This is a clarified version of the article, mentioning that all seasons are referred with respect to northern hemisphere.

Accordingly, the relevant changes are made.

The link to the article in the journal website is https://academic.oup.com/mnras/article-abstract/501/1/1072/5979828.